\documentclass[10pt]{article}
\usepackage{placeins}
\usepackage[utf8]{inputenc}
\usepackage{amsmath}
\usepackage[english]{babel}
\usepackage[margin=1.6cm]{geometry}
\usepackage{graphicx}
\usepackage{booktabs}
\usepackage{caption}
\usepackage{float}
\usepackage[hidelinks]{hyperref}
\usepackage{titlesec}
\usepackage{enumitem}
\usepackage{microtype}
\usepackage{comment} 

\titlespacing*{\section}{0pt}{6pt}{3pt}
\titlespacing*{\subsection}{0pt}{4pt}{2pt}
\setlist{nosep, leftmargin=1.2em}
\newcommand{\yfif}{$y_{50\,\mathrm{ms}}$}
\newcommand{\vyfif}{$v_{y,50\,\mathrm{ms}}$}
\newcommand{\vxfif}{$v_{x,50\,\mathrm{ms}}$}
\newcommand{\dxfif}{$\Delta x_{50\,\mathrm{ms}}$}
\newcommand{\ayearly}{$a_{y,30\text{-}40\,\mathrm{ms}}$}
\newcommand{\aylate}{$a_{y,40\text{-}50\,\mathrm{ms}}$}

\title{\vspace{-1cm}\large Prediction of the maximum penetration of a circular intruder in a two-dimensional granular bed \\
from its early trajectory using Machine Learning.}
\author{\large P. Altshuler \\[2pt]
\normalsize Center for Complex Systems, Physics Faculty, University of Havana, 10400 La Habana, Cuba \\[0pt]
(\normalsize altshulerpatricia@gmail.com)}
\date{\small\today}
\begin{document}
\twocolumn[%
  \maketitle
  \vspace{0.005cm} 
]

\section{Introduction}
The strong nonlinearity of granular media implies the emergence of many phenomena that are difficult to predict on the basis of simple knowledge of the mechanical properties of each grain \cite{Jaeger1996, Martinez2007, Altshuler2008}. Some attempts have been recently made to apply machine learning tools to do the job \cite{Wang2025,Wautier2025}. In particular, the penetration of solid intruders into granular media is a topic with many open questions and potential applications where machine learning could make a difference \cite{Sanchez2014, Katsuragi2007, Diaz2020, Espinosa2023}. Due to its impact on geophysical and industrial phenomena, it is desirable to possess protocols capable of predicting final stages of penetration from early parameters of this process. In this project, 100 falls of a circular intruder onto a bed of expanded polystyrene (EPS) grains are simulated by means of Discrete Element Method (DEM), varying only the intruder density $\rho_{\text{intr}}$. The question addressed is whether, by training a Machine Learning (ML) model, the maximum penetration depth can be predicted using only the early kinematics of the intruder, without providing the model with the value of $\rho_{\text{intr}}$. This amounts to investigating whether the observable dynamics in the first instants already implicitly encodes the mass or density information that determines the maximum penetration for an invariant granular bed.

\section{Description of the data}
The dataset is generated with a DEM simulation (LAMMPS, 2D) \cite{LAMMPS} organized in four stages. Each stage is described below, in execution order.

\begin{description}
\item[\textbf{1. Bed settling}]
Physically, this stage corresponds to filling the container with the granular medium, which is a process with high hysteresis (although we will not analyze it here). It builds a quasi-2D box of $L_x=0.10$~m $\times$ $L_y=0.23$~m. It randomly inserts $\sim$2700 EPS grains (type 2, $\rho_{\text{grain}}=30$~kg/m$^3$) with diameter $d$ sampled uniformly with $\pm10\%$ around a mean diameter of 2~mm. This polydispersity is deliberate. In test simulations the medium slowed the intruder down strangely fast. It was not until several simulations later that the effect of unintentional crystallization of the bed on the intruder became evident. A bed of monodisperse disks tends to crystallize (order itself in a honeycomb-like packing), which would introduce privileged directions in the bed's response to impact; a $\pm10\%$ variation in $d$ suffices to keep the bed disordered without altering its macroscopic properties. The grains fall under gravity and dissipate energy (Hooke contact + restitution damping) until $KE_{\text{total}}<10^{-9}$~J (``settled'' bed criterion). 

\item[\textbf{2. Intruder drop}]
It inserts a circular intruder (radius $0.01$~m) at a fixed position: $x_0=L_x/2=0.05$~m, $y_0=0.116388$~m ($\sim$1~mm above the bed surface, $y_{\text{sup}}=0.105388$~m). The intruder density $\rho_{\text{intr}}$ is received as a command-line parameter, this being the only difference between simulations. The simulation runs in blocks of 50000 steps until the mean velocity of the intruder in a block falls below $0.01$~m/s (intruder ``slowed down''), or up to a safety cap of 60 blocks. Every 500 steps a row is stored with (\texttt{step}, \texttt{time}, $x_{\text{intr}}$, $y_{\text{intr}}$, $v_{x,\text{intr}}$, $v_{y,\text{intr}}$, $\omega_{\text{intr}}$).

\item[\textbf{3. Batch of 100 simulations}] \hspace{0pt}\newline
It samples 100 values of $\rho_{\text{intr}}\sim U[10,45]$~kg/m$^3$, and executes the simulation once per value.

\item[\textbf{4. Extraction of \textbf{\textit{features}} and label}]
For each simulation, all the points of the
trajectory are stored and labelled. In particular,
it stores the maximum penetration, and the 6 features seen in Table 1 of the early penetration (i.e. at the 50 ms instant).
\end{description}

\section{Methodology}
\label{sec:metodologia}

\subsection{Definition of the early window and extraction of \textit{features}}
It was observed that at 20~ms of fall, the trajectories of the 100 simulations are indistinguishable: $x$, $y$, $v_x$, $v_y$ are practically identical, evidencing a free fall without interaction with the bed. Contact with the granular surface occurs at $26.16\pm0.43$~ms in the 100 simulations. Therefore, the extraction window was redefined to $T_{\text{MAX}}=50$~ms. The justification for this specific value is presented in the Supplementary Information at the end of the report. The 6 \textit{features} (Table~\ref{tab:features}) are calculated over the early window $[0,T_{\text{MAX}}]$ with $T_{\text{MAX}}=50$~ms.

\subsection{Exploratory Data Analysis (EDA) with visualizations}
\begin{figure}[H]
\centering
\includegraphics[width=0.48\textwidth]{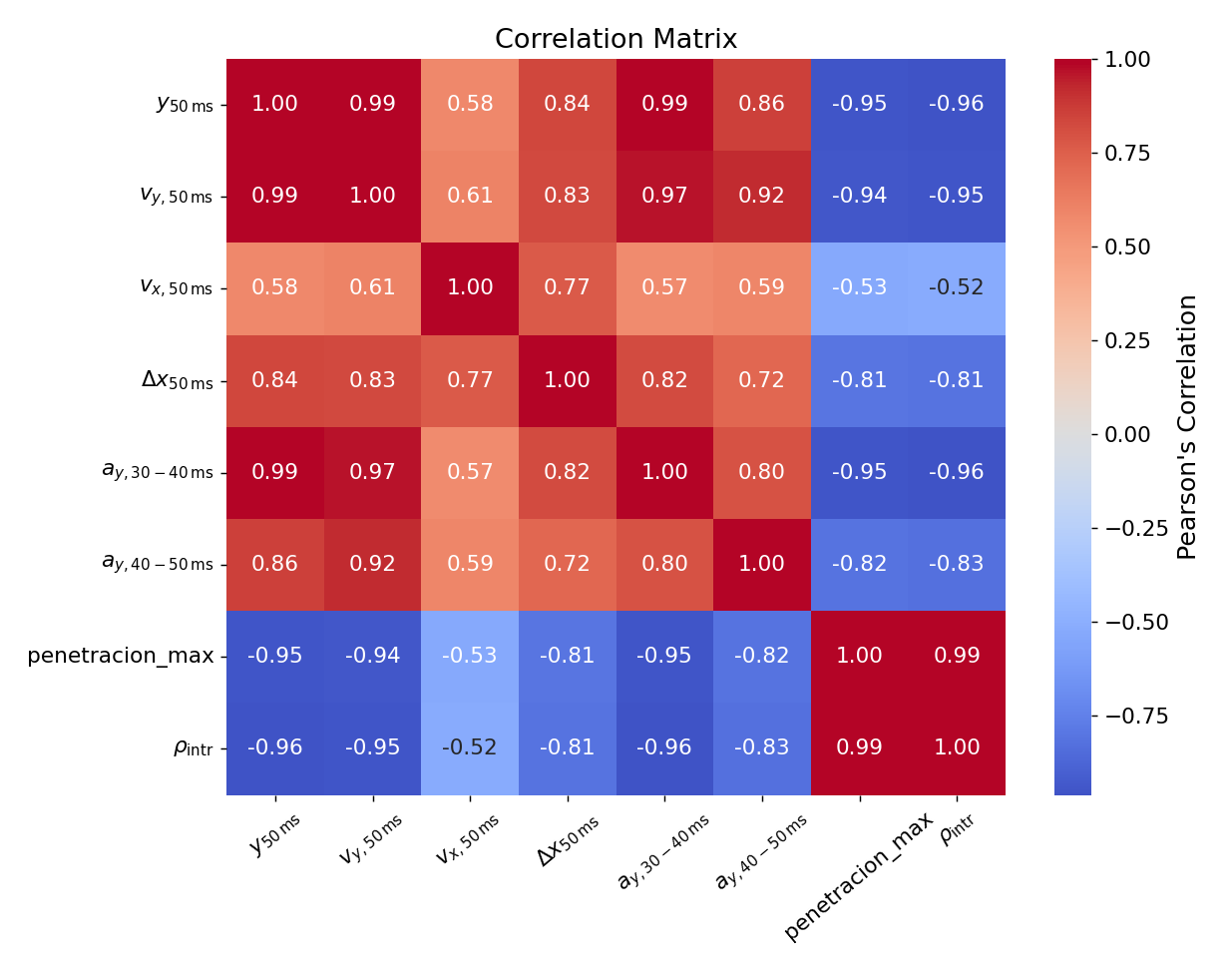}
\caption{Pearson correlation among the 6 \textit{features} (0--50~ms), the label $d_{\text{pen}}$ (\texttt{penetracion\_max}) and the reference intruder density $\rho_{\text{intr}}$ (not used as input). \yfif{}, \vyfif{}, \ayearly{} and \aylate{} show $|r|>0.8$ with the label and among themselves (collinearity).}
\label{fig:correlacion}
\end{figure}

The absence of null values was verified (0 in the 100 rows $\times$ 6 \textit{features}) and the Pearson correlation matrix was calculated (Figure~\ref{fig:correlacion}). The individual correlations with the maximum penetration are high: \yfif{}~$=-0.949$, \ayearly{}~$=-0.948$, \vyfif{}~$=-0.940$, \aylate{}~$=-0.815$, \dxfif{}~$=-0.809$, \vxfif{}~$=-0.530$. There is also high multicollinearity among \yfif{}, \vyfif{}, \ayearly{} and \aylate{} (pairs with $r$ up to $0.99$), so the individual coefficients of a linear model should not be interpreted as independent effects.

\begin{table*}[!htb]
\centering
\footnotesize
\begin{tabular}{@{}llll@{}}
\toprule
\textbf{Feature} & \textbf{Definition} & \textbf{Physical interpretation} \\
\midrule
\yfif{} & intruder height at $t=50$~ms & how much it has sunk by $t=50$~ms (Fig.~\ref{fig:esquema}) \\
\vyfif{} & vertical velocity at $t=50$~ms & sinking speed at $t=50$~ms \\
\vxfif{} & horizontal velocity at $t=50$~ms & residual lateral drift \\
\dxfif{} & $x(50\text{ms})-x(0)$ & accumulated lateral displacement \\
\ayearly{} & slope of $v_y$ in $[30,40]$~ms & bed slowing down shortly after contact \\
\aylate{} & slope of $v_y$ in $[40,50]$~ms & slowing trend at the end of the window \\
\bottomrule
\end{tabular}
\caption{The 6 \textit{features} extracted from the early window $[0,50]$~ms, calculated solely from the intruder kinematics (never from $\rho_{\text{intr}}$).}
\label{tab:features}
\end{table*}

\begin{figure}[H]
\centering
\includegraphics[width=0.5\textwidth]{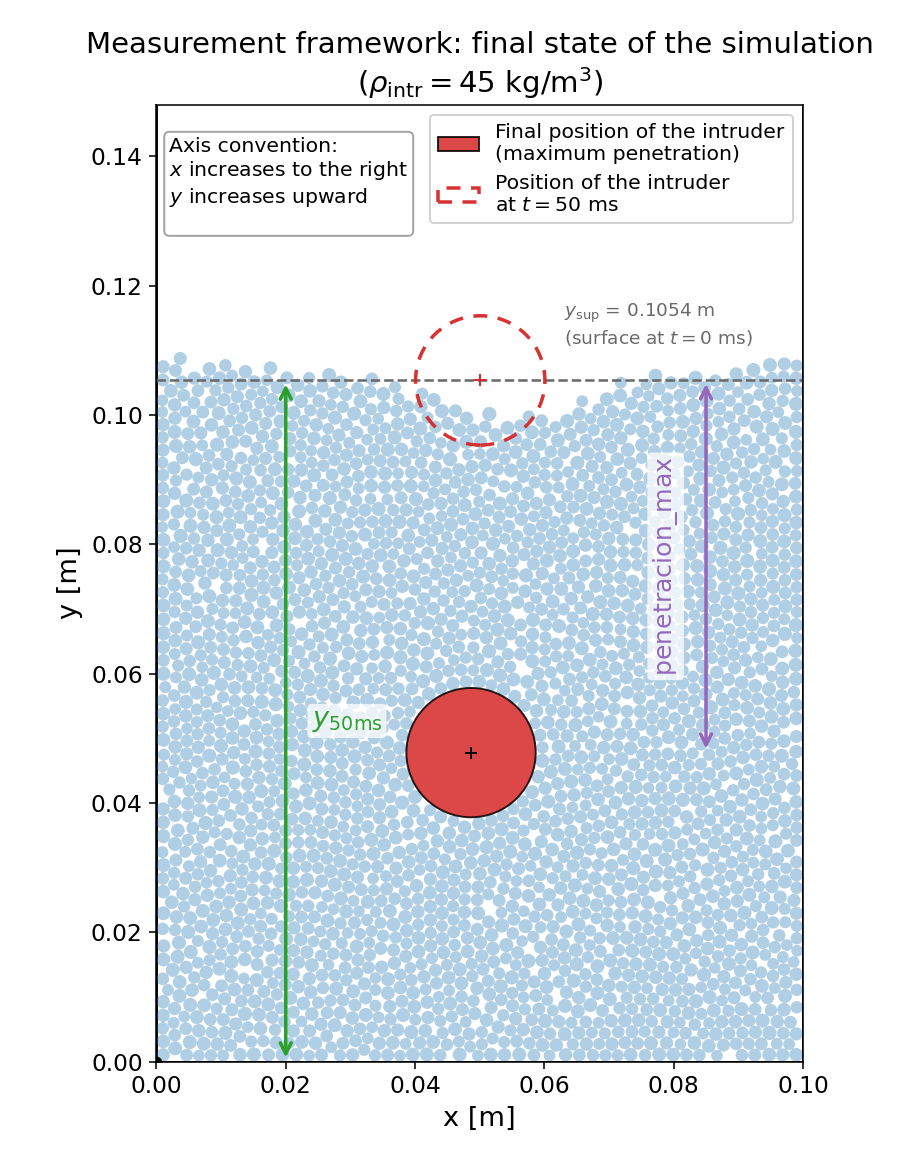}
\caption{Sketch of a typical DEM simulation ($\rho_{\text{intr}}=45$~kg/m$^3$). Axis convention: $x$ increases to the right, $y$ increases upward, origin $(0,0)$ at the lower left corner of the container. \yfif{} (green) is measured from $y=0$ (bottom of the container) to the center of the intruder at $t=50$~ms (dotted red circle). The maximum penetration (purple), is measured from the original bed surface $y_{\text{sup}}$ to the center of the intruder at its position of maximum penetration (solid red circle).}
\label{fig:esquema}
\end{figure}

The negative sign of these correlations is a \textbf{measurement-convention artifact}, not a physically inverse relationship. \yfif{} is measured from the bottom of the container, with $y$ increasing upward (Figure~\ref{fig:esquema}): the more the intruder has sunk by 50~ms, the \emph{smaller} \yfif{} is. In contrast, the maximum penetration is measured from the bed surface downward: the more the intruder sinks, the \emph{larger} the maximum penetration is. Both quantities describe ``how much the intruder sinks,'' but with opposite origins and measurement directions; a larger sinking simultaneously produces a small \yfif{} and a large maximum penetration, hence $r<0$. Physically, \yfif{} and the maximum penetration are \textbf{highly correlated in a positive sense} in terms of ``magnitude of sinking.''

\vspace{1\baselineskip}

\begin{figure*}[!htb]
\centering
\includegraphics[width=\textwidth]{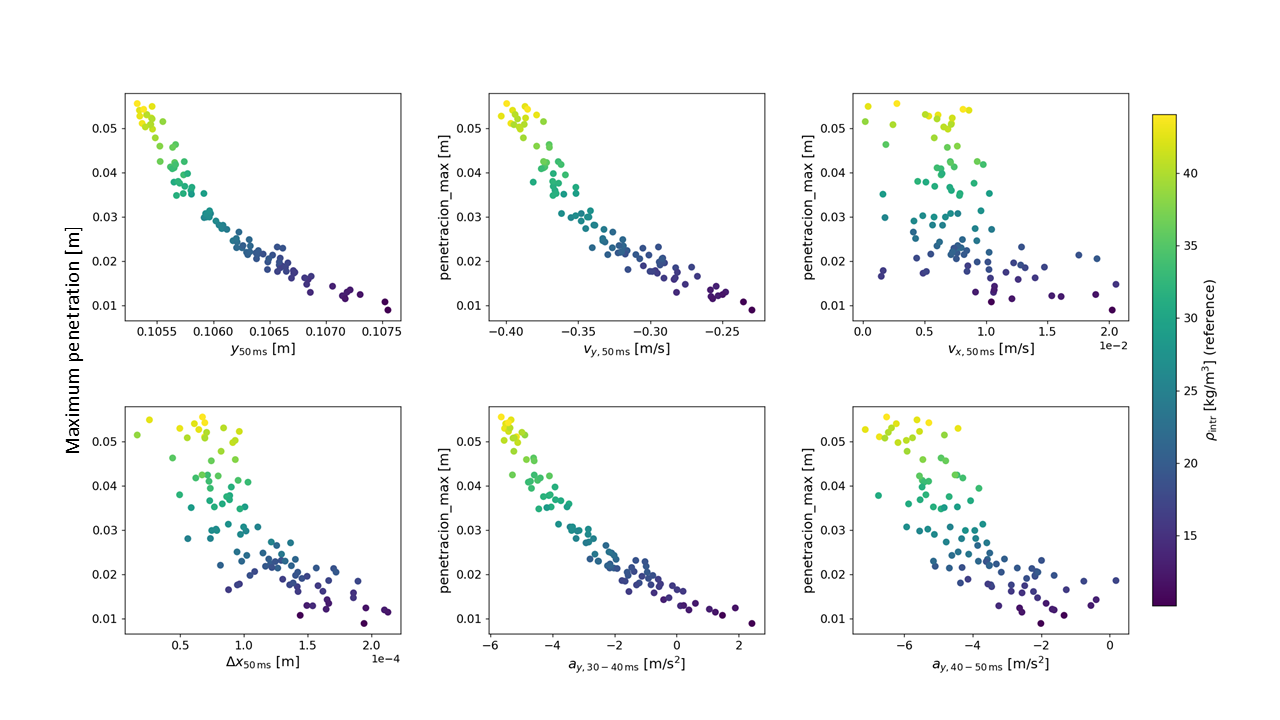}
\caption{Each of the 6 \textit{features} (0--50~ms) vs.\ maximum penetration, colored by $\rho_{\text{intr}}$. The clear color gradient along each trend shows that the kinematic \textit{features} reproduce the ordering by intruder density without having received it as data.}
\label{fig:scatter}
\end{figure*}

The discussed anticorrelation is clearly visible in Figure~\ref{fig:scatter}: \yfif, \vyfif{} and \ayearly{} vs.\ maximum penetration show clean, ordered trends --- the color gradient by $\rho_{\text{intr}}$ progresses monotonically along the trend ---, consistent with $|r|\geq0.94$. In contrast, \vxfif{} and \dxfif{} show scattered point clouds with no clear color pattern. This makes physical sense: lateral displacement is practically negligible in an almost vertical drop, so these two \textit{features} provide little information about $\rho_{\text{intr}}$ and, by extension, about the maximum penetration.

\subsection{Preprocessing}
An 80/20 split was applied (80 \textit{train} / 20 \textit{test}, fixed seed 12345) and a \texttt{StandardScaler} was fitted \textbf{only on the training data} (without information leakage from the test set), then applied to \textit{train} and \textit{test}.

\subsection{Training and evaluation of two models: cross-validation and hyperparameter search}
Two \textbf{regression} models were trained and compared: \textbf{Ridge} (regularized linear regression, interpretable \textit{baseline} with direct coefficients on standardized \textit{features}) and \textbf{Random Forest} (captures nonlinearities and interactions). For each model 80 training samples were used (64 training / 16 validation per iteration) to select hyperparameters, then evaluated on the 20 \textit{test} samples, never seen during fitting. Both models were chosen because the univariate correlations already suggest a strong linear component ($|r|$ up to $0.95$): Ridge serves as a solid baseline and, at the same time, as a reference to detect nonlinearity if Random Forest surpasses it.

\section{Results: plots and evaluation \\
metrics}

\subsection{Model performance}
\begin{figure}[H]
\centering
\includegraphics[width=0.5\textwidth]{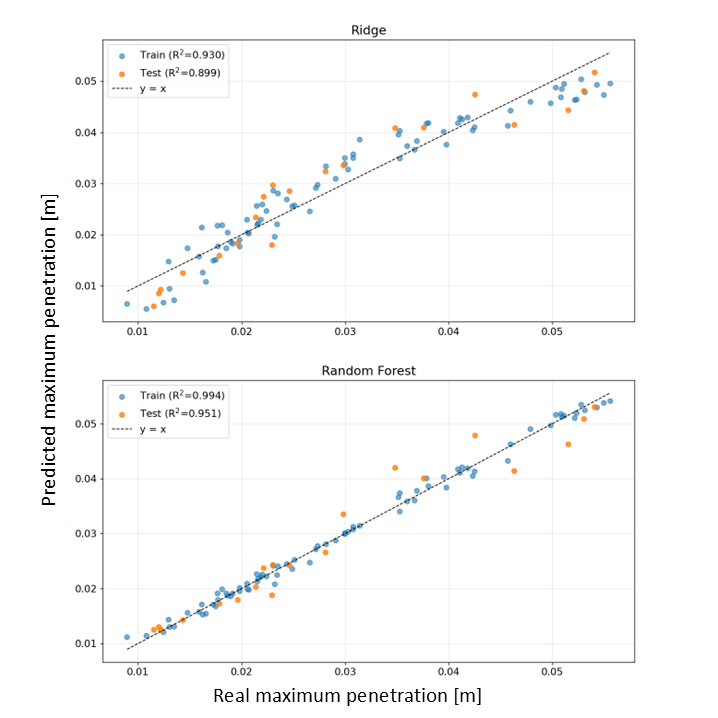}
\caption{Predicted vs.\ real penetration (m) for Ridge (top) and Random Forest (bottom): \textit{train} (blue), \textit{test} (orange) and diagonal $y=x$. Random Forest fits \textit{train} almost perfectly ($R^2=0.994$) and generalizes well to \textit{test} ($R^2=0.951$); Ridge shows greater dispersion at low penetrations ($\sim$10--20~mm).}
\label{fig:predvsreal}
\end{figure}

Table~\ref{tab:resultados} summarizes the performance on the \textit{test} set (20 simulations). Figure~\ref{fig:predvsreal} shows the predicted vs.\ real maximum penetration for \textit{train data} and \textit{test data} in both models. The upper panel corresponds to the Ridge model and the lower one to Random Forest. The diagonal line $y = x$ represents the points where $R^2=1$, that is, where the maximum penetration prediction would be perfect.
\vspace{0.05cm}

Comparing the Ridge model with Random Forest, it is evident that the best-performing model is Random Forest. The \textit{train data} and \textit{test data} points are much closer to $y = x$ than in Ridge, where there is greater dispersion for low penetrations. The improvement in the coefficient of determination when going from Ridge ($R^2=0.899$) to Random Forest ($R^2=0.951$) indicates that the ensemble model captures additional nonlinear information that a purely linear model cannot reproduce. This gain is consistent with the nonlinear nature of granular media, where resistance to penetration does not scale linearly with depth or velocity.

\begin{figure}[H]
\hspace*{-1cm}
\centering
\includegraphics[width=0.58\textwidth]{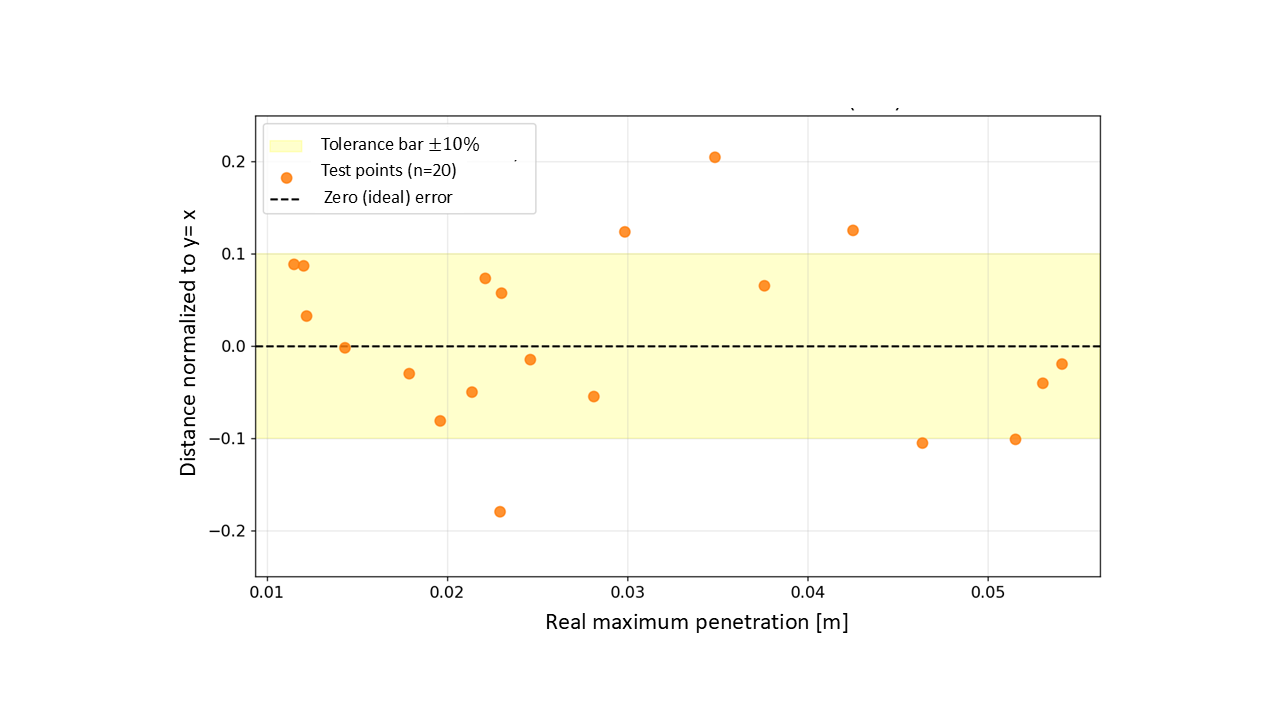}
\caption{Normalized residuals of the Random Forest model on the test set. The $Y$ axis shows the normalized distance to the $y=x$ line, calculated as $(\text{predicted penetration} - \text{real penetration}) / \text{real penetration}$. The yellow band represents the relative error region within the $\pm 10\%$ strip.}
\label{fig:residuos}
\end{figure}

With the aim of understanding in greater depth the predictive power of the Random Forest model, Figure~\ref{fig:residuos} is presented, showing a plot of normalized vertical distance to $y=x$ vs.\ real penetration on the \textit{test data} set trained with this model. This plot provides information on the relative error of the predictions. The yellow band represents the relative error region within the $\pm 10\%$ strip, which is a low percentage error. It is observed that 70\% of the \textit{test data} points (14 of 20) lie within this band, indicating good model performance in terms of relative error for the majority of predictions. \\

\vspace{1\baselineskip}

\FloatBarrier  

\begin{table*}[!htb]
\centering
\small
\begin{tabular}{@{}lccccc@{}}
\toprule
\textbf{Model} & \textbf{Best hyperparameter} & \textbf{CV $R^2$ (train)} & \textbf{Test $R^2$} & \textbf{MAE (mm)} & \textbf{RMSE (mm)} \\
\midrule
Ridge & $\alpha=0.01$ & 0.903 & 0.899 & 4.06 & 4.38 \\
Random Forest & $n{=}300$, depth${=}5$, min\_leaf${=}2$ & 0.973 & 0.951 & 2.32 & 3.05 \\
\bottomrule
\end{tabular}
\caption{Evaluation metrics (regression) for Ridge and Random Forest, with window $T_{\text{MAX}}=50$~ms. CV $R^2$: average of 5-fold cross-validation on the 80 training samples; Test $R^2$, MAE and RMSE: evaluated on the 20 test samples, not used in fitting.}
\label{tab:resultados}
\end{table*}

\FloatBarrier  

The result seems so good that one might question the complexity of the problem posed. The low error of the penetrations predicted by the model, I think, is a direct consequence of using an invariant granular bed for the experiments, which constitutes a limitation of the work since it eliminates the stochastic component inherent to real granular media. This work constitutes a totally valid and interesting problem; however, adding a non-deterministic factor, such as beds settled with independent granular organizations for each simulation, would make the model gain in physical realism, although it would probably lose in precision.


\vspace{1\baselineskip}
\textbf{Which \textit{feature} should be used?}\label{sec:seleccion} Both models were retrained (same split and hyperparameters) with reduced subsets. Using \emph{only} \yfif{} (largest $|r|$ and RF importance), Random Forest goes from $R^2=0.951$ (6 \textit{features}) to $R^2=0.958$, and Ridge from $0.899$ to $0.894$: practically no loss. The subset $\{$\yfif{}, \ayearly{}, \vyfif{}$\}$ (99.4\% of RF importance) gives equivalent results ($R^2=0.895$/$0.954$). Therefore, it is concluded that \yfif{} by itself is, by far, the most useful \textit{feature} for training, while \vxfif{}, \dxfif{} and \aylate{} can be omitted without practical loss of performance.

\section{Physical interpretation of the results and limitations}
\label{sec:limitaciones}
The central result --- $R^2\approx0.90$--$0.95$ without using $\rho_{\text{intr}}$ --- indicates that the information about the intruder density, although never given to the model explicitly, is \textbf{``imprinted'' in the immediate kinematic response to impact}. Because all tests with intruders of various densities were performed on a granular bed with the same configuration, the penetration process becomes deterministic for each intruder. However, the equation describing the penetration of an intruder into a granular medium is a matter of current debate due to the nonlinearity of the forces involved \cite{Katsuragi2007}, so the predictive capacity of the early kinematic parameters and their history (that is, of their filling process, which was identical for the 100 simulations) is surprising. Nevertheless, the forces opposing the penetration process are highly nonlinear \cite{Katsuragi2007}, since they depend on the intermittent appearance and destruction of so-called ``force chains.'' Therefore, we believe that the predictions achieved starting from only the value of the penetration depth at $t=50$~ms have a non-negligible value.
\vspace{0.1cm}

That Ridge ($R^2=0.899$) and Random Forest ($R^2=0.951$) are both high indicates a strong \textbf{linear} component (consistent with $|r|$ up to $0.95$ in EDA) plus an additional real \textbf{nonlinear} component. As discussed qualitatively above, \vxfif{} and \dxfif{} contribute little ($|r|\leq0.81$, importance $\leq0.003$): expected for an almost vertical drop, where penetration is a fundamentally 1D phenomenon (Figure~\ref{fig:scatter}). It is still remarkable that a single parameter at such an early time as 50~ms is sufficient to acceptably predict the maximum penetration. We believe that this situation is also a consequence of using the same settled granular bed for the 100 simulations.

\vspace{0.2cm}
\section{Conclusions}
In this work, we have demonstrated that it is possible to predict with relatively high precision the maximum (final) penetration depth of intruders of various densities in a quasi-2D granular medium, using a simple parameter at early stages of penetration, with Machine Learning. Ridge and Random Forest predict the maximum penetration of the intruder with $R^2\approx0.90$--$0.95$ using kinematic \textit{features} from the first 50~ms of trajectory, never providing the intruder density: the early dynamics already encodes that information. Of the 6 \textit{features}, \yfif{} concentrates almost all the predictive power (Section~\ref{sec:seleccion}): a model trained with only that \textit{feature} matches or surpasses the performance with all 6, so it would suffice to measure the intruder height at $t=50$~ms.

This work can be easily expanded to more realistic situations. For example, using a freshly-constructed granular bed for each simulation; examining intruders with different (non-symmetric) shapes, and penetrations near vertical walls.

\section*{Acknowledgements}
The author thanks valuable suggestions by M. Garc{\'i}a-Borroto and E. Altshuler.

\clearpage
\section*{Supplementary Information}

\subsection*{a) Distribution of \textbf{\textit{features}} and label}
\begin{figure}[H]
\centering
\includegraphics[width=0.5\textwidth]{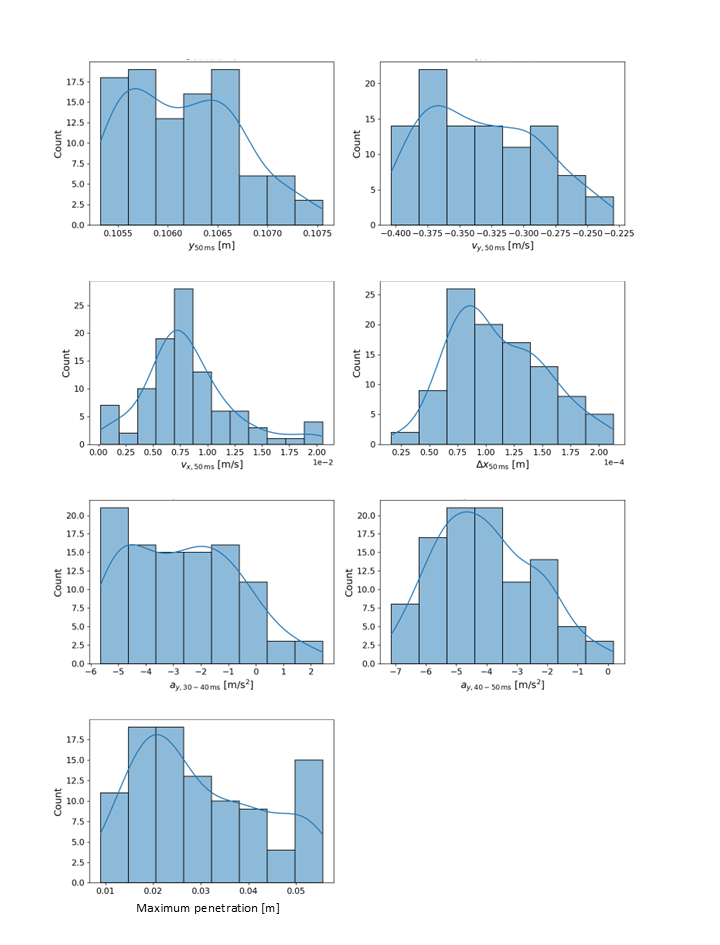}
\caption{Distribution (histogram + KDE) of the 6 examined \textit{features} (0--50~ms) and (\texttt{penetracion\_max}). The label shows two groups: a majority one at 10--30~mm and a smaller one at 50--55~mm; \vxfif{} and \dxfif{} are concentrated and skewed toward small values.}
\label{fig:distribuciones}
\end{figure}

Figure~\ref{fig:distribuciones} shows the histogram with KDE of the 6 \textit{features} and of maximum penetration. \yfif, \vyfif, \ayearly{} and \aylate{} reproduce similar shapes among themselves (consistent with their high mutual correlation and with the label), while \vxfif{} and \dxfif{} concentrate at small values with a long tail.

\subsection*{b) Justification of the window $T_{\text{MAX}} = 50$~ms}
A preliminary analysis showed that at $t=20$~ms (early window originally proposed) \textbf{the 100 simulations are indistinguishable}: $x,y,v_x,v_y$ are identical to the sixth decimal (pure free fall). The instant of first contact, $t_{\text{contact}}$, turned out to be $26.16\pm0.43$~ms (range 23.88--27.36~ms) in the 100 simulations --- determined geometrically, independent of $\rho_{\text{intr}}$. Therefore, the extraction window was redefined to $T_{\text{MAX}}=50$~ms, which guarantees $\geq$22.6~ms of \emph{post-impact} dynamics in the worst case (that is, even for the lightest intruder penetration).

\begin{figure}[H]
\centering
\includegraphics[width=0.5\textwidth]{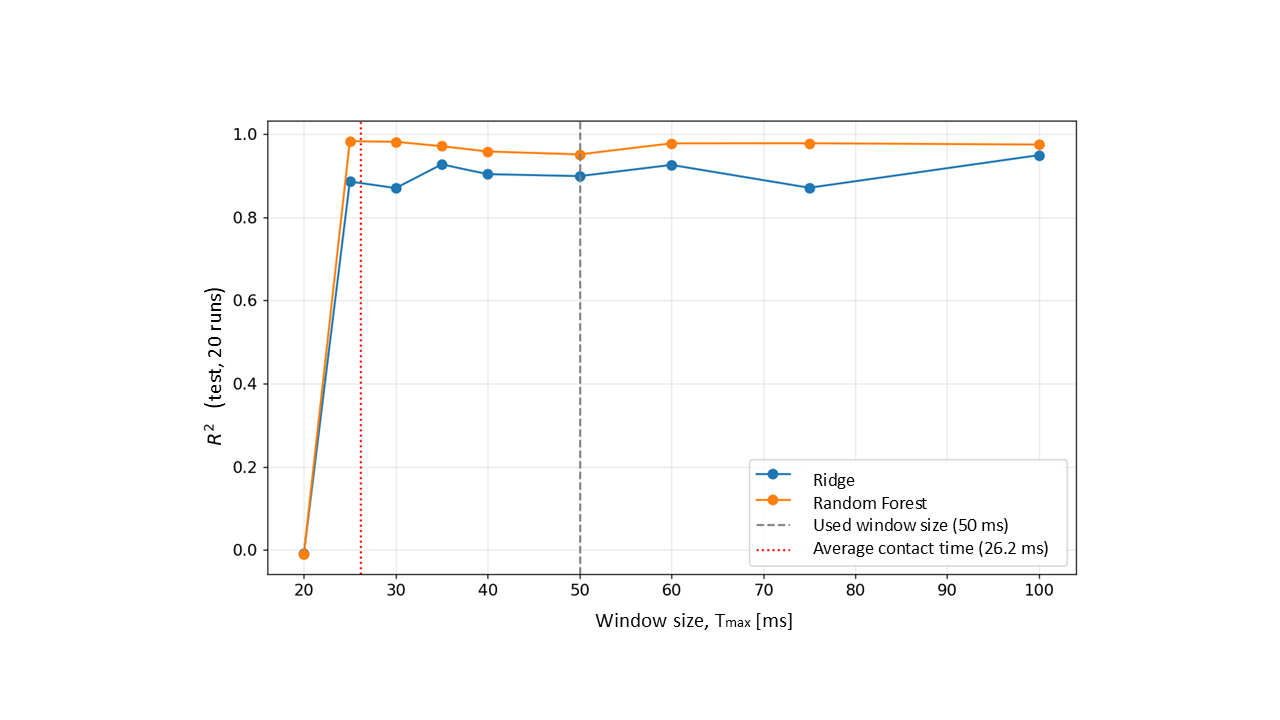}
\caption{$R^2$ (\textit{test}, 20 simulations) of Ridge and Random Forest as a function of the \textit{feature} window size $T_{\text{MAX}}$. The red dotted line marks the average $t_{\text{contact}}$ (26.2~ms); the gray dashed line marks the window used in this report ($T_{\text{MAX}}=50$~ms).}
\label{fig:sweep}
\end{figure}

To confirm that $T_{\text{MAX}}=50$~ms is not an arbitrary value but falls well within the informative regime, we repeated the extraction of the 6 \textit{features} (same definition, generalized to arbitrary $T_{\text{MAX}}$) and the training/evaluation of Ridge and Random Forest for $T_{\text{MAX}}\in\{20,25,30,35,40,50,60,75,100\}$~ms (Figure~\ref{fig:sweep}).

For $T_{\text{MAX}}=20$~ms (before $t_{\text{contact}}$ in all 100 simulations) both models obtain $R^2\approx-0.009$: no predictive power. From $T_{\text{MAX}}=25$~ms (just after $t_{\text{contact}}$) $R^2$ jumps to $\geq0.87$ for both models and remains in that range ($0.87$--$0.98$) for all $T_{\text{MAX}}\geq25$~ms, without a clear monotonic trend. The chosen value, $T_{\text{MAX}}=50$~ms (Ridge $R^2=0.899$, RF $R^2=0.951$), lies comfortably within the informative regime and does not correspond to a local maximum or an isolated minimum: the result is robust to the exact choice of $T_{\text{MAX}}$, as long as it includes a few milliseconds of post-impact dynamics.


\begin{thebibliography}{99}
\bibitem{Jaeger1996}
H. M. Jaeger, S. R. Nagel, R. P. Behringer, \textit{Rev. Mod. Phys.} \textbf{68}, 1259 (1996).

\bibitem{Martinez2007}
E. Mart\'{i}nez, C. P\'{e}rez-Penichet, O. Sotolongo-Costa, O. Ramos, K. J. M\aa l\o y, S. Douady, and E. Altshuler, 
\textit{Phys. Rev. E} \textbf{75}, 031303 (2007).

\bibitem{Altshuler2008}
E. Altshuler, R. Toussaint, E. Mart\'{i}nez, O. Sotolongo-Costa, J. Schmittbuhl, K. J. M\aa l\o y, 
\textit{Phys. Rev. E} \textbf{77}, 031305 (2008).

\bibitem{Wang2025} M. Wang, K. Kumar, Y. T. Feng, {\it et al.}, \textit{Arch. Computat. Methods Eng.} \textbf{32}, 1997 (2025)

\bibitem{Wautier2025} A. Wautier, A. Li, W. Qu, M. Pouragha, F. Nicot, \textit{IOP Conf. Ser.: Earth Environ. Sci.} \textbf{1480}, 012048 (2025)

\bibitem{Sanchez2014}
G. S\'{a}nchez-Colina, L. Alonso-Llanes, E. Martinez, A. J. Batista-Leyva, C. Clement, C. Fliedner,  R. Toussaint, E. Altshuler, \textit{Rev. Sci. Instrum.} \textbf{85}, 126101 (2014).

\bibitem{Katsuragi2007}
H. Katsuragi, D. J. Durian, \textit{Nat. Phys.} \textbf{3}, 420 (2007).

\bibitem{Diaz2020}
V. L. D\'{i}az-Meli\'{a}n, A. Serrano-Mu\~{n}oz, M. Espinosa, L. Alonso-Llanes, G. Viera-L\'{o}pez, E. Altshuler, 
\textit{Phys. Rev. Lett.} \textbf{125}, 078002 (2020).

\bibitem{Espinosa2023}
M. Espinosa, L. Mart\'{i}nez-Ort\'{i}z, L. Alonso-Llanes, L. A. Rodr\'{i}guez-de-Torner, O. Ch\'{a}vez-Linares, 
E. Altshuler, \textit{Sci. Adv.} \textbf{9}, eadf6243 (2023).

\bibitem{LAMMPS}
S. Plimpton, \textit{J. Comput. Phys.} \textbf{117}, 1 (1995). 
https://lammps.sandia.gov

\end{thebibliography}
\end{document}